\documentclass[aps,prl,twocolumn,unsortedaddress,showpacs]{revtex4}

\usepackage{graphicx}
\usepackage{amsmath}
\graphicspath{{.}{./EPS/}}

\newcommand{\lAngle}{\langle \hspace{-0.07cm} \langle}
\newcommand{\rAngle}{\rangle \hspace{-0.07cm} \rangle}   

\begin{document}

\title{Effect of dephasing on 
the current statistics of mesoscopic devices}

\author{Marco G.~Pala$^1$}
\author{Giuseppe Iannaccone$^{1,2}$}
\affiliation{$^1$Dipartimento di Ingegneria dell'Informazione, 
Universit\`a degli Studi di Pisa, via Caruso, I-56122 Pisa, Italy}
\affiliation{$^2$IEIIT, Consiglio Nazionale delle Ricerche,
via Caruso, I-56122 Pisa, Italy}

\begin{abstract}
We investigate the effects of dephasing 
on the current statistics of mesoscopic conductors with a
recently developed statistical model, focusing
in particular on mesoscopic cavities and Aharonov-Bohm rings.
For such devices, we analyze the influence of an arbitrary degree
of decoherence on the cumulants of the current.
We recover known results for the limiting cases of fully coherent 
and totally incoherent transport and are able to obtain
detailed information on the intermediate regime of partial coherence for 
a varying number of open channels.
We show that dephasing affects the average current, shot noise, and
higher order cumulants in a quantitatively and qualitatively
similar way, and that consequently
shot noise or higher order cumulants of the current do not
provide information on decoherence additional or complementary
to what can be already obtained from the average current.
\end{abstract}


\maketitle

Current fluctuations in mesoscopic devices
are due to the randomness of electron transfer \cite{blanter}
and provide detailed information of the underlying transport mechanisms. 
From shot noise measurements  
the distribution functions of open and 
closed transmission channels in mesoscopic samples like 
disordered wires \cite{henny} and the non-integer charge of 
quasi-particles \cite{nic} are extracted.
Such information is not available from conductance measurements
and therefore a more complete description of the
transport mechanisms can be obtained only by computing 
the full-counting statistics of the current \cite{levitov},
which comprises the cumulants of all orders.
Recently, the possibility to explore additional
transport features on cumulants of order higher than the second 
has been experimentally achieved \cite{reulet}.

A highly debated point in these years is 
whether additional precious information on the 
degree of decoherence of a system can be extracted from measurements 
of shot noise or of higher order cumulants of the current.
Several works have addressed this controversial
issue in different structures such as
electronic Mach-Zehnder interferometers \cite{mac1,mac2,mac3}, 
chaotic cavities \cite{cavity} and resonant tunneling diodes 
\cite{aleshkin,reale,butt,chen}.

Indeed, transport in mesoscopic devices is often based on
the coherent evolution of the wave function and is therefore
very sensible to decoherence caused by
the interaction between carriers and the environment \cite{stern}
that reduces the degree of predictability of system evolution. 
Hence, the operation of many mesoscopic devices, especially those based
on quantum interference, can be undermined even by a small degree
of decoherence.
The degree of coherence of transport is typically described
by a unique synthetic parameter, the dephasing length $L_\phi$,
i.e., the characteristic distance over which phase memory is lost.
This fundamental physical quantity
can be experimentally obtained from conductance measurements
on a few suitable devices, 
such as for example Aharonov-Bohm (AB) rings \cite{wl}.

The aim of this letter is to explore the influence of dephasing 
on shot noise and on higher order cumulants of the current
in mesoscopic systems like so called ``chaotic'' cavities and AB rings,
in order to verify which cumulants can be conveniently used
to gain insight into the degree of coherence of transport.

We adopt a recently developed model \cite{pala} that exploits 
the statistical nature of the dephasing process 
and has been recently used to investigate magnetoconductance 
of such structures as a function of the degree of coherence.
The method allows us to introduce an arbitrary degree of
dephasing in the system and to compute all cumulants of the
current, and therefore to investigate their 
dependence on the dephasing length. 
We can anticipate that even for a small number of propagating modes, when
the correspondence principle cannot be invoked, 
{\it essentially all information on 
dephasing can be obtained from conductance properties}.

Let us highlight the fact that, from the modeling point of view, 
researchers typically have to simplify the transport model of a 
mesoscopic conductor, reducing it to the limit of complete
coherence, or to the opposite limit of incoherent transport using
a semiclassical model, which intrinsically rules out interference effects.
Rather intuitively, 
when the Fermi wave length $\lambda_F$ approaches zero or the
number of conducting channels $N_c$ is large, 
approaches based on completely coherent transport should provide the same
result as semiclassical approaches, due to the correspondence principle. 
In such cases, we can expect that the
phase coherence of carriers is irrelevant.
As far as shot noise is concerned, for $N_c \gg 1$,
such behavior is well known, for example, 
for the so-called $1/3$ suppression of shot
noise in diffusive conductors, which has been obtained both with a
quantum mechanical description, 
such as that based on Random Matrix Theory (RMT) \cite{q3}, 
or with statistical simulations \cite{macucci},
and in semiclassical terms, using the
Boltzmann-Langevin equation \cite{s3}. 
A similar agreement has been obtained 
for the so-called $1/4$ suppression of shot noise in
mesoscopic cavities \cite{q4,q44}.
However, shot noise and higher order cumulants of the current in 
the intermediate regime between a fully coherent and incoherent 
transport have not been determined, leaving unsolved
the question of whether they are dependent on the dephasing length.

Ballistic transport in mesoscopic structures is described
within the framework of the Landauer-B\"uttiker theory
\cite{landauer}, which does not allow to include directly the
effects of dephasing. Such effects are
usually treated with phenomenological models, which are based on the
insertion of a virtual voltage probe \cite{buttiker} into the
ballistic region: electrons traveling from source to drain can be
absorbed by a third probe, where they lose their phase
information, and then are re-injected into the conductor.

An alternative phenomenological model which describes decoherence as
a phase randomizing statistical process has been recently proposed
and implemented with a Monte Carlo method by the authors \cite{pala}. 
Such method treats decoherence 
as a random fluctuation of the phase of the propagating modes involved 
in the computation of the scattering matrix (S-matrix), 
and enables us the obtain cumulants of the current from 
Monte Carlo simulations 
over a sufficiently large ensemble of runs \cite{note}.

Current fluctuations in the leads are related to the
number of particles $n$ that traverse the devices during the observation
time $t$. For example,
the first cumulant $\lAngle n\rAngle=\langle I \rangle t/e$ 
gives the mean current $\langle I \rangle$ ($e$ is the absolute value
of the electron charge).
The second cumulant $\lAngle n^2 \rAngle=St/2e^2$ gives
the shot noise power spectral density $S$. Consequently, the Fano factor $F$,
defined as the ratio of $S$ to the power spectral density 
of a Poissonian process
$2 e I$ can be written as the ratio of 
the second to the first cumulant: $F=\lAngle n^2 \rAngle/\lAngle n \rAngle$.
From a numerical point of view, the transmission matrix is 
computed by dividing the domain in $N_x$ slices, then by calculating the 
scattering matrix $s_i$ for each slice,
and by finally composing all the partial S-matrices with the appropriate 
rules \cite{datta} in order to obtain the complete S-matrix of the conductor
$s_T=s_1 \otimes s_2 \otimes \cdots \otimes s_{N_x-1}$.
We include dephasing as a statistical process, 
by adding a random term to the phase 
accumulated by each propagating mode in each slice. Average quantities are
obtained performing simulations on a sufficiently large ensemble of 
runs (of the order of hundreds).
The random term added to each mode of the $i$-th slice 
obeys a Gaussian probability distribution of
zero average and variance $\sigma^2_i$ determined by 
the length of the slice $\Delta x_i$ and by the 
dephasing length $L_\phi$ as 
$\sigma^2_i=\Delta x_i/L_\phi$. By varying $L_\phi$
we are able to explore the complete range of transport regimes, from
completely coherent to completely incoherent.

We consider systems of non interacting electrons at zero temperature, 
and all aspects can be expressed 
in terms of the eigenvalues $T_n$ of the transmission matrix $T=t^\dagger t$. 
Indeed, the $k$-th order cumulants $\lAngle n^k \rAngle$
($k$ is a non-zero integer) of the number of transmitted particles
are defined as the coefficients of the series expansion
of the logarithm of the characteristic function $\chi(\lambda)$ \cite{blanter}:
\begin{equation}
\ln \chi(\lambda)=\sum_k \frac{(i\lambda)^k}{k!} \lAngle n^k \rAngle \;.
\end{equation}

Due to the microscopic mechanism of transport, the 
probability to have $m$ transmitted electrons through a generic channel 
is given by the binomial distribution
$P_m=C_N^m T^m (1-T)^{N-m}$,
where $N$ is the average number of electrons that attempt 
to traverse the device\cite{shimizu,lee}.
The characteristic function is given by 
$\chi(\lambda) = \sum_m P_m \exp(im\lambda) = [T\exp(i\lambda)+1-T]^N$,
which in the case of several independent channels reads
\begin{equation}
\chi(\lambda) = \prod_j [T_j \exp{(i\lambda)} +1 -T_j]^N \;,
\label{chi}
\end{equation}
where the product is performed over all transmission channels in the conductor.
From Eq.~(\ref{chi}) we can obtain the following expression 
for the cumulant of order $k$ \cite{lee}:
\begin{equation}
\lAngle n^k \rAngle =N \sum_j \left. \left[T(1-T)\frac{d}{dT}\right]^{k-1} T 
\right|_{T=T_j} \;.
\label{ce}
\end{equation}

\begin{figure}[ht!]
\includegraphics[width=8cm]{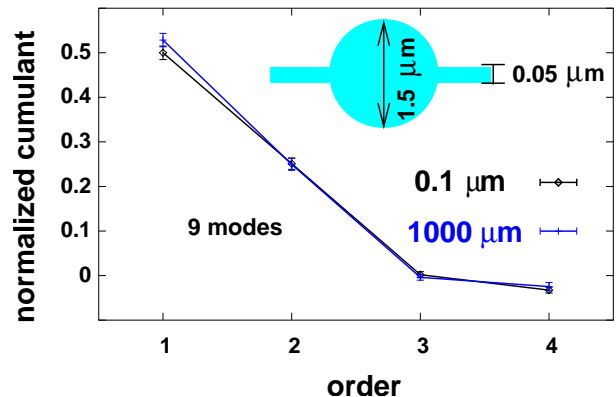}
\caption{(Color online)
Cumulants $\lAngle n^k \rAngle$ for the cavity shown in the inset
for $k=1,2,3,4$, when the Fermi energy allows $N_c = 9$ conducting modes
in the leads.
The two curves correspond to the cases of $L_\phi=0.1$ $\mu$m (incoherent 
transport), and $L_\phi=1000$ $\mu$m (coherent transport).
For $k=2,3,4$ the cumulants are normalized to $\lAngle n \rAngle$.
The first order cumulant is normalized to $N_c$.
The diameter of the circular cavity is 1.5~$\mu$m 
and the lead width is 50~nm.}
\label{fig1}
\end{figure}

As a first case, we focus on a so called ``chaotic'' cavity,
whose structure is shown in the inset of Fig.~\ref{fig1}, and
consisting of a circular cavity with diameter of 1.5~$\mu$m and 
lead width of 50~nm. We consider the material properties of
GaAs. The dwell time of electrons is large enough to provide
the one fourth suppression of shot noise expected for such a
structure \cite{oberholzer}. 
This condition is sufficient
to consider negligible the trajectories responsible 
of back reflection into the lead.

We explore the dependence of the transport properties of the cavity 
on the strength of the decoherence mechanism.
We consider first the two limiting cases of $L_\phi=10^{-1}$~$\mu$m,
which is much smaller than the sample size and therefore corresponds to
an almost fully incoherent regime, and of $L_\phi=10^3$~$\mu$m,
which is much larger than the length 
of the classical path covered by the electrons
inside the cavity, and therefore corresponds to coherent transport.
In Fig.~\ref{fig1} we show that
first four cumulants for $N_c = 9$, obtained from Eq.~(\ref{ce}), are
almost identical in the two cases. 
The total conductance is half the conductance 
of a single constriction, and the Fano factor is $1/4$ in both cases.

\begin{figure}[ht!]
\includegraphics[width=8cm]{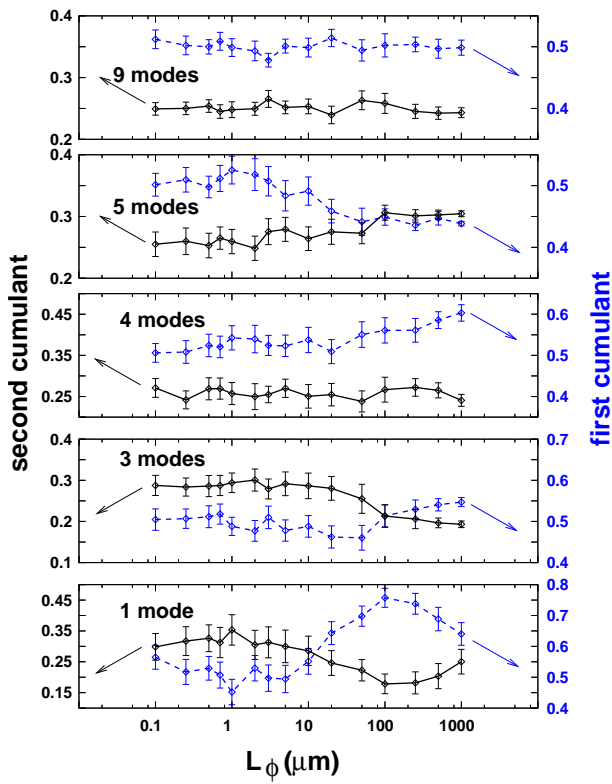}
\caption{(Color online)
Normalized second cumulant $\lAngle n^2 \rAngle/\lAngle n \rAngle$  or 
Fano factor (solid line) of the quantum dot shown in Fig.~\ref{fig1} 
compared with the normalized first cumulant $\lAngle n \rAngle/N_c$ 
(dashed line)
as a function of the dephasing length $L_\phi$. 
From the bottom to the top the Fermi energy allows 
the conduction of 1, 3, 4, 5, and 9 modes.
The error bars in the plot are $\pm 2 s_{\rm m}$,
where $s_{\rm m}$ is the standard deviation of the sample average.
}
\label{fig2}
\end{figure}

\begin{figure}[h!]
\includegraphics[width=8cm]{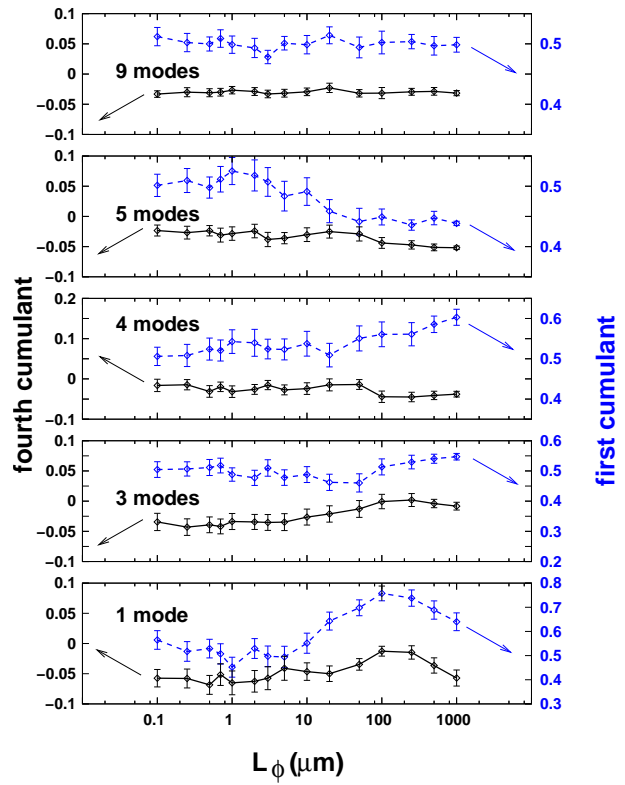}
\caption{(Color online)
Normalized fourth cumulant $\lAngle n^4 \rAngle/\lAngle n \rAngle$ 
(solid line)
compared with normalized first cumulant $\lAngle n \rAngle/N_c$ (dashed line)
of the cavity shown in Fig.~\ref{fig1} as a function 
of the dephasing length $L_\phi$. From the bottom to the top
the Fermi energy allows the conduction of 1, 3, 4, 5, 9 modes, 
respectively.
}
\label{fig3}
\end{figure}

The fact that quantum coherence does not influence the transport
properties for a large number of conducting channels in the structure 
is simply consistent with the correspondence principle.
Hence, we investigated the dependence of the second 
and of the fourth cumulant on the dephasing strength, for 
a smaller number of conducting channels in the leads.

In Fig.~\ref{fig2} and Fig.~\ref{fig3} results are shown
for $N_c= 1, 3, 4, 5, 9$ propagating  modes in the leads,
and a much larger number of corresponding open channels in the cavity.
In such figures we have plotted the first cumulant
of the current $\lAngle n \rAngle/N_c$, 
together with the normalized second cumulant 
$\lAngle n^2 \rAngle/\lAngle n \rAngle$ and the normalized fourth
cumulant $\lAngle n^4 \rAngle/\lAngle n \rAngle$,
as a function of $L_\phi$.

The figures show very clearly that both the second and fourth
order cumulants are largely independent of the dephasing length,
within the range of the error bars due to the finite ensembles considered.
In the case of large $N_c$ the theoretical values 
of both $\lAngle n^2 \rAngle/\lAngle n \rAngle=1/4$
and $\lAngle n^4 \rAngle/\lAngle n \rAngle=-1/32$ are recovered.

In addition, for small $N_c$, whenever a modulation of the 
values of the second or fourth order cumulants can be observed
as a function of $L_\phi$, a corresponding modulation with 
comparable or larger amplitude can be observed in the first cumulant.
In our view this is a demonstration that higher order cumulants
do not provide additional information on dephasing with respect
to that already provided by conductance.
Even in the case of single mode transmission in the lead,
different cumulants are altered by the 
degree of coherence essentially in the same way.

\begin{figure}[h!]
\includegraphics[width=8cm]{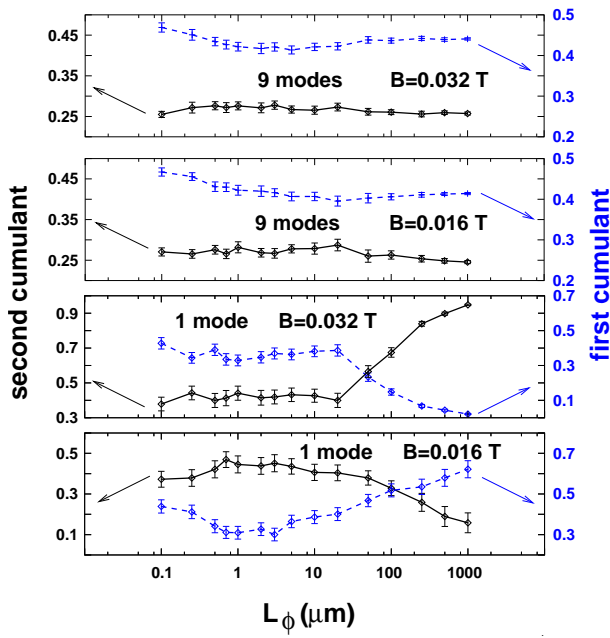}
\caption{(Color online) 
Normalized second cumulant or Fano factor (black solid line)
of the AB ring as a function of the dephasing length
compared with the normalized first cumulant $\lAngle n \rAngle/N$
(blue dashed line). From the top to the bottom we sketch the cases
corresponding to $(N_c=9, B=0.032$~T), $(N_c=9, B=0.016$~T), 
$(N_c=1, B=0.032$ T), and $(N_c=1, B=0.016$~T).
The two values of the magnetic field are such that the magnetoconductance 
presents a maximum (for $B=0.016$~T) and a minimum ($B=0.032$~T)
in the AB oscillations. 
The device is obtained from the cavity sketched in
Fig.~\ref{fig1} by inserting a central
antidot with diameter of 0.9 $\mu$m (not shown).}
\label{fig4}
\end{figure}

The second device we consider is an Aharonov-Bohm ring,
that exhibits very regular magnetoconductance oscillations
due to quantum interference, and is 
therefore very sensitive to the effects of decoherence.
In Fig.~\ref{fig4} we show simulation results for the Fano
factor of an Aharonov-Bohm ring, for the 
cases corresponding to two values of a perpendicular
magnetic field corresponding to a maximum 
($B=0.016$~T) and a minimum ($B=0.032$~T)
of magnetoconductance. 
As can be seen, for the two limiting cases of a small ($N_c=1$)
and a large ($N_c=9$) number of conducting channels,
the Fano factor depends on the 
dephasing length $L_\phi$ in the same way as the conductance,
and therefore does not provide additional insights on decoherence.
 
In this work we have investigated the influence of
dephasing on the transport properties of mesoscopic structures
in order to evaluate the possibility to achieve information on
the degree of decoherence from shot noise properties or
from higher order cumulants of the current. 
We have used a recently developed statistical model to include a 
distributed arbitrary level of dephasing in the device.
We have focused on two types of structures 
with a varying number of propagating channels 
that we believe to be representative of the broad class
of mesoscopic devices.
Our conclusion is that no additional information on the degree
of dephasing is to be expected with respect to that already
provided by the conductance.

The authors would like to thank M.~B\"uttiker for useful discussions
on the model.
Support from the SINANO Network of Excellence 
(EU contract 506844) is gratefully acknowledged.


 

\begin{thebibliography}{99}
\bibitem{note}{With respect to the model presented in Ref.~\cite{pala}, 
here we add a different random term to the phase of the left-going and 
right-going waves of all modes of the scattering matrix. 
In such a way, dephasing affects also weak localization corrections 
to conductance, and reciprocity still holds on the average scattering matrix.}
\end{thebibliography}
\end{document}